\documentclass[11pt,aps,groupedaddress,preprintnumbers,
               nofootinbib]{revtex4}
\usepackage{amsmath,amssymb}
\usepackage{graphicx} %{graphicx,color}
\newcommand{\ket}[1]{\ensuremath{| #1 \rangle}}   % Ket vector
\renewcommand{\vec}[1]{{\mathbf{#1}}}

\begin{document}

\title{On a theory of neutrino oscillations with entanglement}

\author{Boris Kayser}
\email[]{boris@fnal.gov}
\author{Joachim Kopp}
\email[] {jkopp@fnal.gov}
\affiliation
        {\it Theoretical Physics Department, Fermilab, PO Box 500, Batavia, IL 60510, USA}
\author{R.~G.~Hamish Robertson}
\email[]{rghr@uw.edu}
\affiliation{\it Department of Physics and Center for Experimental
        Nuclear Physics and Astrophysics, University of Washington, Seattle, WA 98195, USA}
\author{Petr Vogel}
\email[]{pxv@caltech.edu}
\affiliation
        {\it Kellogg Radiation Laboratory and Physics Department, Caltech, Pasadena, CA 91125, USA}
\date{\today}

\preprint{FERMILAB-PUB-10-194-T}

\begin{abstract}
We show that, despite appearances, a theoretical approach to neutrino
oscillation in which the neutrino and its interaction partners are entangled
yields the standard result for the neutrino oscillation wavelength.  We also
shed some light on the question of why plane-wave approaches to the neutrino
oscillation problem can yield the correct oscillation wavelength even though
they do not explicitly account for the localization of the neutrino source and
the detector.
\end{abstract}

%==============================================================================
\maketitle
%==============================================================================

%------------------------------------------------------------------------------
\section{Introduction}
%------------------------------------------------------------------------------

Is the standard expression \cite{Amsler:2008zz} for the wavelength of neutrino
oscillation correct? Does this expression depend correctly on the underlying
neutrino mass splittings, so that the splittings that have been inferred from
data using this expression are right? 

The quantum-mechanical physics of neutrino oscillation has proved to be quite
subtle, and has been analyzed in a variety of ways over the years. There have
been treatments based on plane waves~\cite{Mann:1976mp, Bilenky:1978nj,
Amsler:2008zz, Lipkin:2005kg}, on neutrino wave packets~\cite{Nussinov:1976uw,
Kayser:1981ye, Giunti:1991ca, Kiers:1995zj, Giunti:2007ry}, and on quantum
field theory~\cite{Grimus:1996av, Grimus:1998uh, Beuthe:2001rc, Giunti:2002xg,
Akhmedov:2010ms}. (A thorough summary has been given recently by Akhmedov and
Smirnov \cite{Akhmedov:2009rb}.)  Most of these treatments have yielded the
standard expression for the probability of oscillation, but the correct way to
derive this expression is still occasionally disputed.  The analysis presented
in the present work, based on plane waves and on an assumed entanglement
between the oscillating neutrino and a recoil particle, finds an oscillation
wavelength that appears to be very different from the usual one, but turns out
to be physically equivalent to it.

In this paper, we recap the recent treatment of oscillation using entangled
plane waves. We express a number of concerns about such an analysis, and about
plane wave treatments in general. However, we also comment on why plane wave
treatments, while not strictly correct and consistent, nevertheless often yield
the correct oscillation wavelength. We demonstrate that, in spite of concerns,
and in spite of the fact that the oscillation wavelength found by considering
entangled plane waves {\em appears} to be markedly different from the standard
one, this wavelength, properly understood, is physically completely equivalent
to the usual one, and yields the same neutrino mass splittings when applied to
data. This result is the main point of this paper.

%------------------------------------------------------------------------------
\section{Plane-wave description with entanglement}
%------------------------------------------------------------------------------

In this section, we present a derivation of the oscillation wavelength,
focusing on neutrinos produced in two-body decays of the form ${\rm P}
\rightarrow \nu + {\rm R}$, in which a parent $P$ decays into a neutrino $\nu$
and a recoil $R$. In an effort to avoid apparent paradoxes that can arise in
standard plane-wave treatments of neutrino oscillations, we assume that the
neutrino and its recoil form a two-body entangled state.  The role of
entanglement in neutrino and neutral kaon oscillations has been considered by
others \cite{Goldman:1996yq, Nauenberg:1998vy, Cohen:2008qb, Dolgov:1997xr,
Burkhardt:2003cz, Burkhardt:1998zj, Lowe:1996ct, Akhmedov:2010ua}.  In the
analysis presented here, the oscillation wavelength that emerges corresponds to
the separation of the neutrino and its associated recoiling particle at a
common time in the parent rest frame.  This wavelength is not the usual
wavelength that would be observed in an experiment (because the recoiling
particle is not usually observed), but we show that the two wavelengths are
physically equivalent and lead to the same neutrino survival or appearance
probability.%
\footnote{This section incorporates arguments presented previously in
unpublished form by one of us (RGHR)~\cite{Robertson:2010xrv1}. A revised
version of that work~\cite{Robertson:2010xr} takes into account insights
explained in the present paper.}

A neutrino of flavor $\alpha$ is a linear combination of mass eigenstates
$\nu_i$ with masses $m_i$,
\begin{eqnarray}
\left | \nu_\alpha \right \rangle = \sum_i U_{\alpha i}^*\left | \nu_i \right \rangle,
\label{eqtwo}
\end{eqnarray}
where the $U_{\alpha i}$ are elements of the Maki-Nakagawa-Sakata-Pontecorvo
mixing matrix \cite{Amsler:2008zz}.   In the approach that assumes
entanglement, energy-momentum conservation requires that the energy and
momentum of the recoil in the decay  ${\rm P} \rightarrow \nu + {\rm R}$ depend
on which $\nu_i$ was actually emitted. In the rest frame of the parent, $p_R
= -p_\nu$, where $p_R$ and $p_\nu$ are the momenta of the recoil
and neutrino, respectively. In terms of the coordinates in that frame,
the two-particle wave function may be written as
\begin{eqnarray}
\left |R{\nu}; x, X \right \rangle
  &\sim& \sum_i U_{\alpha i}^* \ket{R_i(p_{P,i},X)} \ket{\nu_i(p_{\nu,i} x)} \nonumber\\
  &=&    \sum_iU_{\alpha i}^*\left |R\right \rangle\left |{\nu_i}\right \rangle
         e^{-iE_{R,i} t - ip_{\nu,i} X}e^{-i E_{\nu,i} t + ip_{\nu,i} x} \nonumber \\
  &=&    \left |R\right \rangle e^{-iE_at}\sum_iU_{\alpha i}^*\left |{\nu_i}\right \rangle
         e^{ ip_{\nu,i} D }. 
\label{eqfive}
\end{eqnarray}
Here, $\left |R\right \rangle$ describes the internal degrees of freedom of the
recoil, and $\left |{\nu_i}\right \rangle$ simply identifies one of the three
neutrino mass eigenstates. $E_{\nu,i}$ and $E_{R,i}$ are the energies of the
neutrino and recoil, respectively, and $t$ is the time.  The $P$-rest-frame
distance between the neutrino and the recoil is $D \equiv x - X$, where the
coordinates $x$ and $X$ are the positions of the neutrino and recoil,
respectively.  The energy $E_a = E_{\nu,i} + E_{R,i} = m_P$ in the third line
of Eq.~\ref{eqfive} is just the mass of the parent $P$.  The index $i$ on
$p_{\nu,i}$, $E_{\nu,i}$, and $E_{R,i}$ indicates that these quantities all
depend on $i$, while $E_a$, as well as $t$, $x$, and $X$, do not.  (To describe
a neutrino oscillation experiment with a spatial resolution much better than
the oscillation length, we have to fix $(t,x)$ and $(t,X)$ at the
$P$-rest-frame coordinates of the spacetime points where the neutrino and
recoil, respectively, are detected.  These points are defined by the experiment
and therefore do not depend on $i$.) Note that the third line of
Eq.~\ref{eqfive} shows that, in the rest frame of the parent, the two-particle
system consisting of the neutrino and the recoil can be described by a
one-particle wave function depending only on the relative coordinate $D$. This
is analogous to the treatment of the hydrogen atom in most quantum mechanics
textbooks, where the two-body wave function of the electron and the nucleus is
factorized into a one-particle wave function describing the relative motion of
the two particles, and a one-particle wave function describing the center of
mass motion (which is trivial in the center of mass frame).

Suppose that in the laboratory frame, the parent $P$ is moving to the right
along the $x$-axis with speed $\beta$. The lab-frame location $x'$ of the
neutrino when it is at the spacetime point $(t,x)$ is $x'=\gamma(x+\beta t)$.
The corresponding lab-frame location $X'$ of the recoil at the same time $t$ in
the $P$ rest frame is  $X'=\gamma(X+\beta t)$.   Thus, the lab-frame distance
between the neutrino and recoil at time $t$ in the $P$ rest frame is $x'-X' =
\gamma(x-X) \equiv D'$.  Consequently the state $\left|R\nu; x,X\right\rangle$ of
Eq.~\ref{eqfive} may be rewritten as
\begin{eqnarray}
  \left |R{\nu}; D' \right \rangle &= & e^{-i E_at}\left |R\right \rangle
  \sum_iU_{\alpha i}^*\left |{\nu_i}\right \rangle  e^{i \gamma^{-1} p_{\nu,i}D'},
\label{eqtena}
\end{eqnarray}
where $p_{\nu,i}$ is the parent rest-frame momentum of the neutrino when the
latter is $\nu_i$. The leading energy-dependent phase factor 	in
Eq.~\ref{eqtena}  is unobservable.    As a result, the interference effects of
neutrino oscillations arise {\em solely} from the different {\em momenta} in
the components in the final state.   In the oscillation probability at lab-frame
separation $D'$, $P(\nu_\alpha \rightarrow \nu_\beta)= \left|\left \langle R\nu_\beta|
R\nu_\alpha \right \rangle_{D'} \right|^2$, the $i-j$ interference term depends on
$D'$ through the phase factor $ \exp{[i \gamma^{-1}( p_{\nu,i}-p_{\nu,j})D']}$.
Thus, the wavelength $\lambda'_{D,ij}$ of oscillation in the recoil-neutrino
separation $D'$ is determined by 
\begin{eqnarray}
\gamma^{-1}( p_{\nu,i}-p_{\nu,j})\lambda'_{D,ij} & = & 2\pi. 
\label{eqsix}
\end{eqnarray}
Now, to leading order in $\Delta m_{ij}^2 = m_i^2-m_j^2$, 
\begin{eqnarray}
 p_{\nu,i}-p_{\nu,j} & = & - \Delta m_{ij}^2 \frac{1}{2m_P}\frac{m_P^2+m_R^2}{m_P^2-m_R^2},
\label{eqseven}
\end{eqnarray}
where $m_R$ is the mass of the recoil. Thus, apart from an irrelevant sign, 
\begin{eqnarray}
 \lambda'_{D,ij} & \simeq & \frac{4\pi\gamma m_P}{\Delta m_{ij}^2}\frac{m_P^2-m_R^2}{m_P^2+m_R^2}.
\label{eqeight}
\end{eqnarray}

To compare this result with the standard expression for the wavelength of
neutrino oscillation, it is useful to rewrite it in terms of the ``neutrino
beam energy'' $E_0'$, defined as the energy that massless neutrinos would have
in the laboratory. Since $E_0' = \gamma(1+\beta)(m_P^2-m_R^2)/2m_P$, and the
speed $\beta_R$ of the recoil in the $P$ rest frame obeys $1+\beta_R =
2m_P^2/(m_P^2+m_R^2)$, we have
\begin{eqnarray}
 \lambda'_{D,ij} & = & \frac{4\pi E_0'}{\Delta m_{ij}^2}\frac{1+\beta_R}{1+\beta}.
 \label{eqnine}
 \end{eqnarray}
 
Equations \ref{eqeight} and \ref{eqnine} define a wavelength in  laboratory
coordinates for  the separation $D'$ between the neutrino and the recoil.  In
the standard expression \cite{Amsler:2008zz} for neutrino oscillation, the
survival probability oscillates as a function of the lab-frame distance $L'$
between the neutrino source and the detector with a wavelength $
\lambda'_{L,ij}$ given by
\begin{eqnarray}
 \lambda'_{L,ij} & = & \frac{4\pi E_0'}{\Delta m_{ij}^2}.
 \label{eqninea}
 \end{eqnarray}

We will show in the next section that these two expressions, Eqs. \ref{eqnine}
and \ref{eqninea}, in fact give equivalent results for the laboratory
wavelength of neutrino oscillations measured in the standard fashion by
detecting neutrinos at a known distance from a source region.

A comment is in order about the role that the entanglement between the neutrino
and the recoil plays for the oscillation phenomenology. In particular, one may
wonder if the oscillation pattern is modified if the recoil undergoes an
interaction that breaks the entanglement long before the neutrino is detected.
In fact, in a typical experiment, the recoil will interact with the matter that
makes up the neutrino source very soon after it has been produced. If it is
unstable, it might also decay very rapidly.  However, it is easy to see that
such interaction or decay cannot change the neutrino oscillation phenomenology.
First, the order in which the recoil interaction or decay and the neutrino
detection occur depends on the Lorentz frame in which we are working. The
neutrino flavor transition probabilities, on the other hand, are Lorentz
invariant, so they cannot depend on the time ordering of these processes. In a
more formal way, the same conclusion can be reached by using the quantum
amplitude approach to particle mixing probabilities developed
in~\cite{Kayser:1995bw}.  This approach, which takes the entanglement between
the neutrino and the recoil into account and confirms the validity of
Eq.~\ref{eqfive}, yields the joint probability for the recoil to interact (or
decay) at one spacetime point, and the neutrino to produce a charged lepton of
a given flavor $\alpha$ at another spacetime point. To determine the
consequences of this approach for neutrino oscillations, one must integrate
this joint probability over all possible interaction points of the recoil. But
it turns out that the joint probability depends on the recoil interaction point
$(T, X)$ only through a phase factor $e^{-i E_{R,i} T - i p_{\nu,i} X}$,
analogous to the recoil phase factor we have encountered in the second line of
Eq.~\ref{eqfive}. It is easy to show that this factor is independent of which
neutrino mass eigenstate has been emitted together with the
recoil~\cite{Smirnov:2010, Akhmedov:2010ua}. Thus, it is an overall phase that
does not influence the quantum interference between neutrino mass eigenstates
that leads to neutrino oscillations. Hence, the oscillation pattern is
independent of where or how soon the recoil interacts or decays, destroying its
entanglement with the neutrino.

%------------------------------------------------------------------------------
\section{Physical equivalence of  wavelengths}
%------------------------------------------------------------------------------

Since the wavelengths $ \lambda'_{D,ij}$ and $ \lambda'_{L,ij}$ in
Eqs.~\ref{eqnine} and \ref{eqninea} describe oscillations in two different
variables, $D'$ and $L'$, let us see how these variables are related.  We take
the decay  ${\rm P} \rightarrow \nu + {\rm R}$ to occur at the spacetime point
$(t,x) = (t',x') = (0,0)$.  Suppose that the neutrino is then detected at a
subsequent time $t$ in the $P$ rest-frame.  Then,  in that frame,   if the
neutrino's speed is  $\beta_\nu$,   it will have traveled a distance
$L=\beta_\nu t$ to its point of detection.  The $P$ rest-frame separation $D$
between the neutrino and the recoil when the neutrino is detected will be $D =
\beta_\nu t + \beta_R t$.  Thus, since the neutrino is ultrarelativistic, with
$\beta_\nu \simeq 1$, 
\begin{eqnarray}
\frac{D}{L} & = & 1 + \beta_R.
\label{eqnineb}
\end{eqnarray}
Now, in the laboratory frame, the distance $L'$ between the neutrino source
(i.e., the location of $P$ when the decay ${\rm P} \rightarrow \nu + {\rm R}$
occurred) and the point of detection is $L'=\gamma(L + \beta t)\simeq
\gamma(1+\beta)L$, where we have once again used $\beta_\nu \simeq 1$.
Furthermore, as we have already seen, at the neutrino detection time $t$ in the
$P$ rest-frame, the laboratory-frame distance $D'$ between the neutrino and the
recoil is related to its $P$ rest-frame counterpart $D$ by
$D'=x'-X'=\gamma(x-X)=\gamma D$.  Thus, the variables $D'$ and $L'$ to which
the wavelengths $ \lambda'_{D,ij}$ and $ \lambda'_{L,ij}$ refer are related by 
\begin{eqnarray}
\frac{D'}{L'} & = & \frac{\gamma D}{\gamma(1+\beta)L} = \frac{1+\beta_R}{1+\beta}.
\label{eqninec}
\end{eqnarray}
By comparison, from Eqs.~\ref{eqnine} and \ref{eqninea},
\begin{eqnarray}
\frac{ \lambda'_{D,ij}}{ \lambda'_{L,ij}}&=& \frac{1+\beta_R}{1+\beta}.
\label{eqnined}
\end{eqnarray}
That is, the ratio between the ``new'' wavelength  $\lambda'_{D,ij}$  of
oscillation in $D'$, and the standard wavelength $\lambda'_{L,ij}$ of
oscillation in $L'$, is exactly the same as that between $D'$ and $L'$.  

Thus, although the new  $\lambda'_{D,ij}$ of Eq.~\ref{eqnine} has emerged from
an approach that entails entanglement, while the standard $\lambda'_{L,ij}$ of
Eq.~\ref{eqninea} has come from analyses that generally do not, these two
wavelengths differ only because they refer to two different, alternative
distance variables.  Thus, they are physically equivalent. Properly used to fit
given oscillation data, they would yield precisely the same neutrino
squared-mass splitting $\Delta m_{ij}^2$. 

In practice, of course, an analysis of experimental data using Eq.~\ref{eqnine}
instead of the usual Eq.~\ref{eqninea} would require knowledge of the
coordinates of both the neutrino and the recoil at the same time $t$ in the
parent rest frame. The required coordinates of the recoil would be extremely
difficult to obtain. However, it is not necessary to actually detect the recoil
if the spacetime trajectory and energy of the parent particle are known---as
they would be, for instance, in an electron-capture beta-beam neutrino
experiment. In that case, energy-momentum conservation could be used to infer
the trajectory of the recoil from the kinematics of the parent and the measured
coordinates of the neutrino.

%------------------------------------------------------------------------------
\section{Why is the same $\Delta m^2$ obtained?}    
%------------------------------------------------------------------------------

So long as the standard approaches and the approach that invokes entanglement
are all valid ways of deriving the neutrino oscillation wavelength, the
wavelengths derived by these approaches obviously must be correct and
physically equivalent. In the previous section, we have demonstrated the
physical equivalence for a particular approach that invokes entanglement and
that obtains the wavelength for oscillation in the separation $D'$ between the
neutrino and its recoiling partner at a given time $t$ in their common parent's
rest frame. But this must be a more general result. Consider a ``gedanken''
neutrino oscillation experiment, with some relevant properly defined distance
$D''$ between the neutrino and its recoiling partners, and with $D''$
proportional to the usual source-to-detector distance $L'$. Then, the ratio
between the wavelengths of oscillation in $D''$ and $L'$ will be $D''/L'$, so
that these two wavelengths will be physically equivalent.

Another way of understanding intuitively why the standard expression for the
oscillation length is obtained even when the neutrino and the recoiling
particle are considered as an entangled state is the following: Let us choose a
frame in which the component of the recoil that is entangled with neutrino mass
eigenstate $\nu_1$ of mass $m_1$ is at rest; i.e.\ $p_{R,1}'' = 0$. We denote
kinematic quantities in this frame by a double prime. From simple kinematic
arguments, it follows that
\begin{eqnarray}  
  p_{R,j}'' &=& \frac{\Delta m_{j1}^2 \, m_R}{m_P^2 - m_R^2} + \dots \\  
  E_{R,j}'' &=& m_R + \dots \,, 
\end{eqnarray} 
where `\dots' denotes terms that are at least 4-th order in the neutrino
masses, divided by combinations of $m_R$ and $m_P$. The complex phase of the
$j$th component of the entangled state has the form
\begin{equation}  
  -E_{R,j}'' t'' + p_{R,j}'' X''  - 
   E_{\nu,j}'' t'' + p_{\nu,j}'' x'' \,.
\end{equation}
We see that, at leading order in $\Delta m_{j1}^2$, the first term is a
constant that does not contribute to the phase differences relevant to neutrino
oscillations and can therefore be omitted. We will now show that the second
term can be neglected as well.  Since we are using a plane wave approach here,
it is not immediately obvious what $X''$ is; after all, plane waves are
delocalized over space.  We will argue below that it is reasonable to impose
the relations $x'' = v_\nu'' t''$ and $X'' = v_R'' t''$ by hand, where it is
justified to take $v_\nu''$ and $v_R''$ to be the averages of the group
velocities associated with the individual components of the neutrino and the
recoil, respectively.  In the frame where $p_{R,1}'' = 0$, all $v_{R j}''$ are
proportional to $\Delta m_{j1}^2$, so that no matter how exactly the average
$v_R''$ is defined, $p_{R,j}'' X''$ is second order in $\Delta m_{j1}^2$ and
therefore negligible.  Consequently, the phase reduces to the standard
expression
\begin{equation}  
  -E_{\nu,j}'' t'' + p_{\nu,j}'' x'' 
\end{equation}
that does \emph{not} depend on the properties of the recoil. Due to Lorentz
invariance, the phase must therefore be the standard one in any frame.

%------------------------------------------------------------------------------
\section{Plane waves vs.\ wave packets}
%------------------------------------------------------------------------------

It is clear that, due to the Heisenberg uncertainty principle, neutrinos
produced and detected in localized regions of spacetime must have a nonzero
spread in energy and momentum. Many authors have argued that, to take this
spread into account, a fully consistent theoretical treatment of neutrino
oscillations requires wave packets~\cite{Nussinov:1976uw,Kayser:1981ye,
Giunti:1991ca,Grimus:1996av,Beuthe:2001rc,Akhmedov:2010ms}. Other authors have
argued that wave packets are unnecessary~\cite{Stodolsky:1998tc}.

Indeed, in most neutrino oscillation experiments, wave packet effects, such as
wave packet separation due to different group
velocities~\cite{Nussinov:1976uw}, are negligible, and simplified plane wave
approaches can correctly predict experimental results.  The deeper reason for
this can be understood if we note that the coordinate space representation of a
typical Gaussian neutrino wave packet,
\begin{eqnarray}
  \psi_j(\vec{x}, t) & \propto & e^{-i E_{j0} t + i \vec{p}_{j0} \vec{x}}
    \exp\bigg[ \frac{(\vec{x} - \vec{v}_j t)^2}{4 \sigma_x^2} \bigg] \,,
  \label{eq:wp}
\end{eqnarray}
is simply a plane wave, multiplied with an enveloping Gaussian.  Here, the
index $j$ distinguishes different neutrino mass eigenstates, $\vec{p}_{j0}$ and
$E_{j0}$ are the average momentum around which the wave packet's momentum
distribution is peaked and the associated energy; $\vec{v}_j =
\vec{p}_{j0}/E_{j0}$ is the group velocity, and $\sigma_x$ is the width of the
wave packet in coordinate space.  The neutrino oscillation probability depends
on the phase differences between wave packets associated with different mass
eigenstates $j$.  We observe that, at each fixed spacetime point, these phase
differences depend only on $E_{j0}$ and $\vec{p}_{j0}$.%
\footnote{Note that interference can only occur between wave packet components
located at the \emph{same} point in spacetime.}
They do \emph{not} depend on $\sigma_x$, so they are independent of the precise
shape of the wave packets, and, in particular, remain unchanged in the limit
$\sigma_x \to \infty$ corresponding to plane waves. They do, however, vary over
space and time. In the full wave packet picture, the enveloping Gaussian
ensures that only space-time points along the trajectory $\vec{x} \sim \vec{v}
t$, and located at the detector site, $\vec{x} \simeq \vec{L}$, contribute to
the oscillation probability.%
\footnote{Here, $\vec{v}$ should be understood as an average of the individual
$\vec{v}_j$, which is a valid concept as long as the wave packets have not yet
separated. Also note that $|\vec{v}| \simeq c$, with small corrections to this
relation being a negligible second-order effect in the small neutrino masses.}
In a plane wave approach, this has to be ensured by imposing
$\vec{x} = \vec{v} t$ and $\vec{x} = \vec{L}$ by hand.

Another argument not to invoke wave packets is the observation by
Kiers, Nussinov, and Weiss~\cite{Kiers:1995zj} that a continuous flux of
neutrino wave packets with identical momentum distributions cannot be
distinguished from an ensemble of plane wave neutrinos, whose individual
momenta follow the same momentum distribution. In fact, the density matrices
describing the two ensembles are identical.

\begin{figure}
  \begin{center}
    \includegraphics[width=7cm]{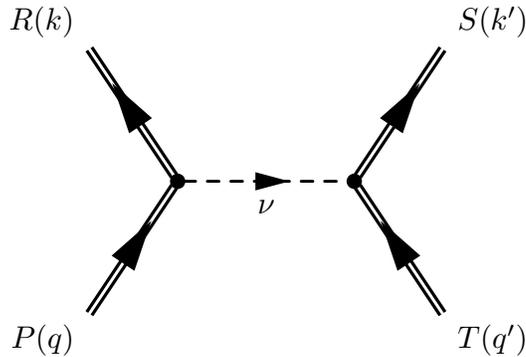}
  \end{center}
  \caption{The Feynman diagram of neutrino oscillations. $P$ and $R$
    are the parent particle and the recoil, respectively, while $T$
    is the target atom at the detector and $S$ denotes the outgoing
    ``signal'' particles.}
  \label{fig:feyn}
\end{figure}

Even though the above arguments show that plane wave approaches to neutrino
oscillations can be justified, there are inconsistencies with the quantum field
theoretic (QFT) formalism, where the production and detection processes are
explicitly included in the calculation and the neutrino is treated as an
internal line in a Feynman diagram (see Fig.~\ref{fig:feyn}).  Since energy and
momentum are exactly conserved at the production and detection vertices, the
intermediate neutrino can only be an on-shell energy-momentum eigenstate (as it
has to be if it propagates over macroscopic distances) if its interaction
partners---the external lines of the Feynman diagram---are also energy-momentum
eigenstates. However, it is clear that the external energies and momenta in
this case cannot be the same for diagrams involving different neutrino mass
eigenstates. Since in quantum mechanics interference is only possible between
amplitudes for processes that describe different paths leading to the
\emph{same} final state, this means that no oscillations are
possible~\cite{Beuthe:2001rc}.

A similar argument can also be invoked to show that, in the QFT treatment,
entanglement between the neutrino and its interaction partners cannot persist
asymptotically for $t \to \infty$.  Such asymptotic entanglement would lead to
different external states in diagrams involving different neutrino mass
eigenstates, so interference between these diagrams would be impossible. If we
take the point of view that initially the neutrino is entangled with at least
some of its interaction partners, we are led to the conclusion that these
particles must be disentangled by further interactions for oscillations to take
place. (This has been noticed previously in ref.~\cite{Cohen:2008qb}.)
Disentanglement is achieved when the particles interact with their environment.
These interactions localize them and thus introduce momentum uncertainties that
are usually large enough to allow external particle states entangled with
different neutrino mass eigenstates to interfere in spite of their different
energies and momenta.

With this in mind, one can justify a plane wave approach even in QFT.  The
procedure is to compute the amplitudes corresponding to different neutrino mass
eigenstates and then sum them \emph{coherently} (before squaring), keeping in
mind that such coherent summation is only justified if the final state
particles undergo further interactions that introduce energy and momentum
uncertainties larger than the energy and momentum differences between the
different entangled states.

If one wishes to avoid such reasoning, one can also directly incorporate the
effect of the localizing secondary interactions into the Feynman diagram
computation by treating the external particles as wave packets with appropriate
energy and momentum spreads.  Different neutrino mass eigenstates will then
simply couple to different (though usually overlapping) portions of the
external wave packets' momentum distributions.  Feynman diagrams involving
different neutrino mass eigenstates can thus have identical external states, so
that interference and therefore neutrino oscillations emerge.

%------------------------------------------------------------------------------
\section{Conclusions}
%------------------------------------------------------------------------------

For the reasons we have explained, one must be cautious when following a plane
wave approach to neutrino oscillation, whether or not the approach incorporates
entanglement. A plane wave approach does not include the neutrino source and
detector localizations that are physically essential if oscillation with
distance is to be observed. Consequently, such an approach cannot be used to
discover all the implications of these localizations.  However, the explicit
inclusion of localization effects, which is automatic in wave packet
treatments, can be replaced for some purposes by physical reasoning in a plane
wave approach.  Indeed, the oscillation wavelength derived from a plane wave
analysis, with or without entanglement, can be perfectly correct. In
particular, as we have shown, the wavelength that emerges from the analysis
presented in this paper, which does invoke entanglement, is physically
equivalent to the (correct) wavelength obtained by standard wave packet or
plane wave treatments. Thus, one may continue to have confidence in the
neutrino mass-squared splittings $\Delta m^2$ that have been deduced by
applying the standard wavelength expression to data.

%------------------------------------------------------------------------------
\section*{Acknowledgments}
%------------------------------------------------------------------------------

We would like to thank S.~Parke for organizing a very fruitful discussion
session about entanglement in neutrino oscillations at Fermilab. We are also
indebted to E.~Akhmedov, J.~Conrad, G.~Garvey, T.~Goldman, M.~Goodman,
B.~Keister, J.~Lowe, M.~Messier, M.~Shaevitz, A.~Smirnov, R.~Volkas, W.~Winter,
and L.~Wolfenstein for inspiring and useful discussions.  The work of PV was
partially supported by the US Department of Energy under Contract No.\
DE-FG02-88ER40397.  The work of RGHR was supported by the US Department of
Energy under Contract No.\ DE-FG02-97ER41020. Fermilab is operated by Fermi
Research Alliance, LLC under Contract No.\ DE-AC02-07CH11359 with the US
Department of Energy.

% Bibliography

\end{document}